\documentclass[twocolumn,superscriptaddress,aps,reprint]{revtex4-2} % revtex 4-2 is now the standard (~Oct. 2020)
\usepackage{lmodern}
\usepackage[T1]{fontenc}
\usepackage{exscale,amsmath,amsbsy, amssymb} % The exscale option is included to be compatible with lmodern's math fonts
\usepackage{bm}% bold math (vectors)
\usepackage{xspace} % needed for math spacing issues
\usepackage{comment}   
\usepackage{graphicx, xcolor, url}
\usepackage{mathtools}
\usepackage{enumitem}
\usepackage{siunitx}	% SI units
\usepackage{textcomp} %% so that microtype doesn't get mad
\usepackage{microtype}

\usepackage{hyperref} % always put hyperref last
\hypersetup{colorlinks}
\newcommand{\smod}{\mu} % other choice: G
\newcommand{\stress}{\sigma} % stress tensor
\newcommand{\stension}{\gamma} % surface tension
\newcommand{\Cph}{C_{\text{p}}}  % phase velocity (non-dim)
\newcommand{\Cg}{C_{\text{g}}}  % group velocity (non-dim)
\newcommand{\Fr}{\mathrm{Fr}} % Froude number
\newcommand{\Bo}{\mathrm{Bo}} % Bond number = \sqrt(rho g d^2/gamma)
\newcommand{\lce}{\ell_\text{ce}}   % elastic–capillary length
\newcommand{\leg}{\ell_\text{eg}}   %  elastic–gravity length
\newcommand*{\vc}[1]{\bm{#1}}   % vectors
\newcommand{\dif}{\mathop{\mathrm{{}d}}\mathopen{}} % differential
\newcommand{\tpartderiv}[2]{\partial{#1}/{\partial{#2}}}		% inline partial derivative
\newcommand{\nfpartderiv}[3]{\frac{\partial^{#1}{#2}}{\partial{#3}^{#1}}}	% general display partial derivative
\def\coleq{\coloneqq} % definition (left *defined by* right)
\newcommand{\FT}[1]{\tilde{#1}} % Fourier transform
\renewcommand{\Re}{\operatorname{Re}}   % real part
\newcommand*{\khat}{\hat{\vc{k}}}

\newcommand{\vdel}{\bm{\nabla}} % gradient operator

\newcommand{\lap}{\nabla^2} % laplacian

\newcommand{\ci}{\mathrm{i}}    % imaginary number
\newcommand{\ee}{\mathrm{e}}    % Euler's number
\renewcommand{\leq}{\leqslant}

\begin{document}
\title{Surface wakes on ultra-soft solids}

\author{Aditi Chakrabarti}
%\thanks{A.C. and D.J. contributed equally to this work}
\altaffiliation{A.C. is currently at Schlumberger-Doll Research, Cambridge, MA 02139, USA.}
\affiliation{School of Engineering and Applied Sciences, Harvard University, Cambridge, MA 02138, USA}

\author{Divya Jaganathan}
\thanks{A.C. and D.J. contributed equally to this work.}
\affiliation{School of Engineering and Applied Sciences, Harvard University, Cambridge, MA 02138, USA}

\author{Robert Haussman}
\affiliation{School of Engineering and Applied Sciences, Harvard University, Cambridge, MA 02138, USA}

\author{L.\ Mahadevan}
\thanks{Corresponding author: \href{mailto:lmahadev@g.harvard.edu}{lmahadev@g.harvard.edu}}
\affiliation{School of Engineering and Applied Sciences, Harvard University, Cambridge, MA 02138, USA}
\affiliation{Departments of Physics and Organismic and Evolutionary Biology, Harvard University, Cambridge, MA 02138, USA}

\date{\today}

\begin{abstract}
We explore the dynamical response of the free surface of an ultra-soft solid driven by a localized moving pressure disturbance. Experiments reveal a steady V-shaped wake analogous to a surface Mach wedge. A simple geometric argument provides a qualitative explanation consistent with observations. A theoretical framework combining elastodynamic, capillary, and gravitational effects yields a generalized dispersion relation that smoothly interpolates between Kelvin’s theory of liquid interface wakes and Rayleigh’s theory of elastic surface waves. Our analysis explains the observed Mach-like behavior quantitatively while also emphasizing how elastodynamic effects can generate effective damping through radiative leakage. Together, our experiments and theory reveal the existence of a “soft wake” regime that bridges fluid and solid surface-wave physics, offering new routes for probing the dynamics of soft surfaces.  
\end{abstract} 

\maketitle

Soft elastic solids are easily deformable when subjected to external loads, and are characterized by relatively small (zero-frequency) shear moduli. The interface of a soft elastic solid is even softer than the bulk solid because in addition to a weak material response, it is soft by virtue of geometry (i.e., it is free on one side).  Thus, soft interfaces are susceptible to both body forces (e.g., gravity) and interfacial forces (e.g., surface tension). For a solid with density $\rho$, elastic shear modulus $\mu$, a surface energy per unit area $\gamma$, subject to a gravitational body force per unit volume $\rho g$, one can define two natural length scales: an elastocapillary length,  $\ell_{ce}=\gamma/\mu$  and an elastogravity length, $\ell_{eg} = \mu/\rho g$. For ultra-soft solids, elastic, capillary and gravitational effects all become simultaneously important when the elastocapillary length and the elastogravity length become comparable. Then, letting $\ell_{eg} \sim \ell_{ce}$ yields an expression for the shear modulus $\mu^* \sim (\rho g \gamma)^{1/2}$; substituting $\rho=1000$ kg/m$^3$, $g = 10 m/s^2$, $\gamma=10 mN/m$ yields $\mu^* \sim 10 Pa$. The recent ability to create very weak materials \cite{Aditi_gelsurfacetension} with ultra-low moduli in this range, i.e., $\mu ~\sim O(10)  Pa$ makes this capillary-gravity-elasticity regime accessible via tabletop experiments~\cite{chakrabarti2016elastobuoyant,ReviewImagingTissue_Greenleaf,capElast2020,Dufresne2025}.  

However, little is still known about the dynamics of ultra soft solids, when there is a delicate interplay between inertia, elasticity, gravity and capillarity, e.g., when a soft interface is driven by a transient pulse  \cite{choi1992measurements,onodera1998surface,shao2018extracting}, a topic that has implications for fields such as soft elastography \cite{bercoff2004supersonic,jacob2007nonlinear} and non-destructive imaging~\cite{chakrabarti2014vibrations, harden1991hydrodynamic}.  Here, we study the 2-dimensional wake left by a localized pressure source moving on the surface of a 3-dimensional solid, using experiment and theory to understand the wake on an ultra-soft solid interface, a problem connecting two classical problems: Rayleigh waves on the surface of an elastic  solid~\cite{rayleigh1885waves} and the Kelvin wake behind a ship moving in deep water~\cite{thomson1887ship}. 

\begin{figure}[t!]
\centering
\includegraphics[trim={2cm, 0.5cm, 0.5cm, 0.4cm}, clip, width=.92\columnwidth]{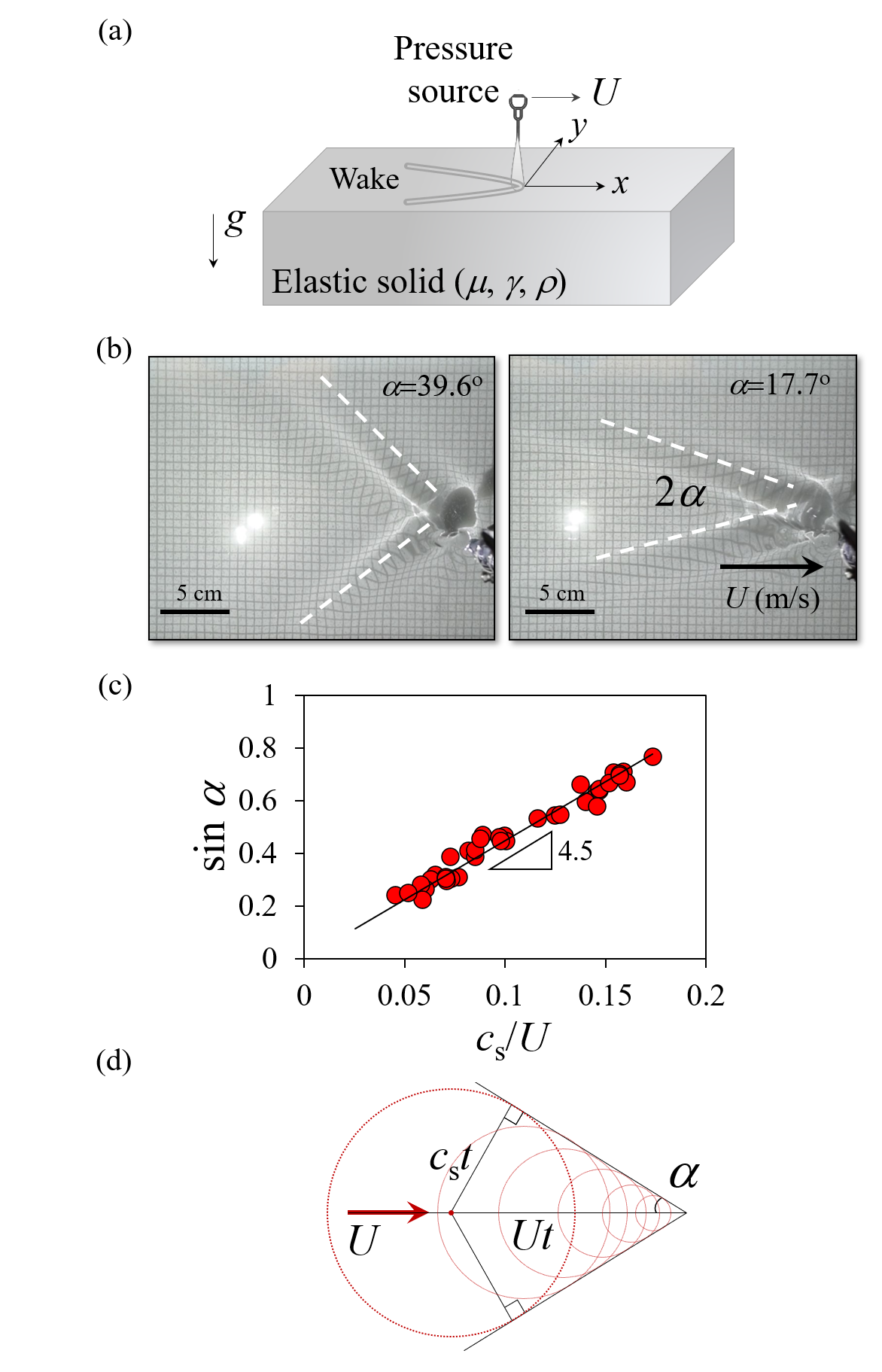}
\caption{(a) Schematic of the experimental setup, where a pressure source of compressed air moving at speed $U$ impinges on the surface of an elastic solid, generating a surface wake. (b) Top-view images of wakes for two representative speeds, $U=0.55 ~m/s$ (left) and $U=1.28~ m/s$ (right) leads to different opening angles ($2\alpha$) on an ultra-soft solid ($\mu \approx 6.5 Pa$). The wake angle decreases with the increasing speed of the pressure source. A projected grid on the surface of the gel slab from the top enables visualization of the surface deformation. (c) Experimental data showing Mach-like scaling, where $\sin\alpha$ varies linearly with $c_s/U$, and $c_s = \sqrt{\mu/\rho}$ is the Rayleigh (shear) wave speed. (d) Schematic of the geometry of wave fronts on a soft solid surface due to a moving pressure source.}
\label{experimentandData}
\end{figure}

To create a  dynamic  pressure source moving at a speed $U$ above a soft elastic solid (Fig.~\ref{experimentandData}a), we start with a large tank filled with about 10 litres of an ultrasoft polyacrylamide hydrogel with elastic shear modulus, $\mu \approx {6.5Pa}$ (See SI  for details of the protocol to create and measure the rheology of the gel) and density $\rho=1000$kg/m$^3$. The dimensions of the soft solid slab are 50 cm $\times$ 25 cm with a depth $H=$ 8 cm. The pressure source was created by forcing compressed air out through a thin nozzle ($d=$ 0.84mm), mounted on a rail that can slide horizontally above the gel surface at a constant height of $\sim 4$cm above it. As the pressure source moves at a uniform speed ($U$) above the gel surface, we observe a wake behind the moving disturbance (Fig.~\ref{experimentandData}b, SI movie S1,S2), with ridges of maximum surface elevation on the soft interface forming a linear wedge-like profile.  The opening angle of the leading dominant ridge ($2 \alpha$) depends on the speed of the pressure source $U$; faster sources yield wakes with smaller angles. Indeed, plotting $\sin{\alpha}$ versus the scaled inverse speed of the pressure source $c_s/U$, where $c_s=\sqrt{\mu/\rho}$ is the shear-wave speed in the soft solid, we see a linear relationship with $\sin \alpha \approx 4.5 c_s/U$ (Fig.~\ref{experimentandData}c). 

A natural comparison of this observation is with the wake observed on liquid surfaces in the `deep-water' limit~\citep{thomson1887ship,Lamb1932,lighthill1978waves,rabaud2013ship,pethiyagoda2021}. However, the wake angle here deviates substantially from the Kelvin angle (i.e, 19.4$^\circ$) characteristic of gravity wakes at low Froude numbers ($\Fr < 1$) in a few ways. Firstly, the wake is sharper and more V-like without the feathered decorations seen in gravity waves. Secondly, the angle  varies with the speed of the source and does not saturate to the Kelvin limit for classical gravity-capillary wakes   \cite{rabaud2014gravcap}. Finally, in contrast to gravity–capillary wakes, the wake patterns we observe exhibit only faint fine features away from the source, indicating strong damping (See SI movie S1 for a comparison between soft gel and water wakes). 

To get a qualitative understanding of our observations, we first consider a minimal model for the elastic wake, ignoring, in the first instance, the effects of gravity and capillarity. On the surface of an incompressible solid, freely propagating non-dispersive Rayleigh waves propagate with a speed proportional to the shear wave speed $c_s$ \cite{rayleigh1885waves}.  If a pressure source moves on the surface with a speed $U$ greater than the $c_s$, we expect a Mach-like wake to emerge.  A geometric argument for the wake half-angle, $\alpha$, follows by considering the superposition of the circular wavefronts (Fig.~\ref{experimentandData}d) and leads to the simple relationship, 
\begin{equation}\label{Mach}
    \sin{\alpha}=A c_s/U, 
\end{equation} 
which is qualitatively consistent with  our experimental measurements where $A\sim 4.5$, i.e., the wake angle increases with decreasing speed. To explain the weakly dispersive nature of the wake (see SI movie 1,2), and the prefactor requires a quantitative theory of surface waves including  the contributions from gravity, surface tension and elastodynamics that we now turn to. 

We assume that the free surface bounds a soft incompressible elastic half-space  ($z\leq 0$) of uniform density $\rho$ and shear modulus $\smod$. Displacements of the solid are given by $\vc{\xi}(\vc{x},z,t)\equiv(\vc{u},w)$, where $\vc{x}=(x,y)$.  In the linear elastodynamic regime relevant for our observations, the dynamical equations of motion read
\begin{equation}\label{eq:EOMdispl}
    \rho\nfpartderiv{2}{\vc{\xi}}{t} = -\vdel P + \smod \lap\vc{\xi},
\end{equation}
 subject to the condition of incompressibility $\nabla \cdot \vc \xi = 0$, which upon substituting into eq.~(\ref{eq:EOMdispl}) gives  
\begin{equation}\label{eq:EOMP}
    \lap P = 0.
\end{equation}
Since we are interested in the response to small-amplitude disturbances at the surface, we take $P$ to be the excess pressure from the pre-stressed equilibrium configuration, so that the boundary conditions deep in the interior are given by $ \text{$\vc{\xi}\to\vc{0}$ and $P\to 0$ as $z\to -\infty$,}$. On the surface, capillary forces provide a stress proportional to the surface's mean curvature, yielding the free-surface boundary conditions $ \stress_{iz,0} = \smod (\partial w/\partial x_i + \partial u_i/\partial z )_{z=0} = 0,~    \stress_{zz,0} = (-P + 2\smod~\partial w /\partial z )_{z=0} = (P_z + \stension\lap_\parallel w - \rho g w)_{z=0},$
where $i = x, y$ and $\vdel_\parallel = (\tpartderiv{}{x},\tpartderiv{}{y})$, and $P_z(\bm{x},t;U,L)$ is the external moving pressure source applied onto the surface, moving in the $x$ direction with a constant speed $U$. For simplicity, we assume an isotropic pressure source with a characteristic size $L$.

We solve the linear eqs.~(\ref{eq:EOMdispl}) and (\ref{eq:EOMP}) subject to the boundary conditions using 2-dimensional Fourier methods, assuming sufficient far-field decay, using the standard definition $f(\vc{x},z,t) = \int\dif{\omega}\int\frac{\dif^2{\bm{k}}}{(2\pi)^2}\FT{f}(\vc{k},z,\omega)\,\ee^{\ci(\vc{k}\cdot\vc{x} - \omega t)}$, where $\tilde{f}$ is its corresponding transform, and the arbitrary function $f$ can be $P$, $\vc{u}$, and $w$. This yields the complete solution (see section B1 in SI for details)
\begin{align}
    \FT{P} &= \frac{\omega^2(\omega^2-2c_s^2k^2)\,\ee^{k z}}{D(k,\omega)}\FT{P}_z(k,\omega), \\
    \FT{\vc{u}} &= (\ci\khat)\frac{(\omega^2-2c_s^2 k^2) k\,\ee^{kz} + 2(k^2 c_s^2)q\,\ee^{qz}}{\rho D(k,\omega)}\,\FT{P}_z(k,\omega), \\
    \FT{w} &= \frac{(\omega^2-2c_s^2 k^2)k\,\ee^{kz} + 2(k^2 c_s^2)k\,\ee^{qz}}{\rho D(k,\omega)}\,\FT{P}_z(k,\omega), \label{eq:ftw_expr}
\end{align}
where $k = |\bm{k}|$, $q:=\sqrt{k^2 - (\omega/c_s)^2}$, and the denominator:
\begin{equation}\label{Dexpr}
\begin{aligned}
    D(k,\omega) = \omega^2\Bigl(g k + \frac{\gamma}{\rho}k^3\Bigr) &- (2c_s^2 k^2-\omega^2)^2 \\
    &+ 4 c_s^3 k^3\sqrt{c_s^2  k^2 -\omega^2} ~,
\end{aligned}
\end{equation}
which is consistent with the characteristic equation derived in \cite{shao2018extracting} for the 1-dimensional case.

For a disturbance pressure source moving with uniform velocity, as in our experiment, the Doppler shift sets $\omega= \bm{U} \cdot \bm{k}$. The stationary surface displacement field is then obtained from the Fourier transform of eq.~(\ref{eq:ftw_expr}) evaluated at $z=0$ as follows \cite{havelock1908propagation},
\begin{equation}
    w(\bm{x}) =  \frac{1}{(2\pi)^2 \rho} \int \int d^2 \bm{k} \: e^{i \bm{k} \cdot \bm{x}}\frac{(\bm{U}\cdot \bm{k})^2 k}{D(k,\bm{U}\cdot \bm{k})}\tilde{P}_z(k;L) ~, \label{eq:surface_elev_integral}
\end{equation}
where $\bm{x} - \bm{U}t \rightarrow \bm{x}$, with $\bm{U}=U \bm{e}_x$, is the new coordinate vector in the moving frame. The stationarity condition,
\begin{equation}\label{eq:CoS}
    \omega = \bm{U} \cdot \bm{k} = Uk\cos\phi~, 
\end{equation}
where $\phi$ is the angle between the wave vector $\bm{k}$ and the direction of moving source (see Fig.~\ref{fig:EM-A}c), restricts the surface waves that interfere constructively to those with phase speed $c_p(k) \coleq \omega/k < U$. Furthermore, the dispersion relation for surface waves is given by $D(k,\omega)=0$. We now examine the expression for $D$ to understand the effects of elasticity.

To work with dimensionless quantities, we use the pressure source size $L$ as the reference length and $U/L$ as the reference frequency, and define $K\coleq k L$  and $\Omega \coleq \omega L/U $. This induces the transformation $(L^3/U^2 g)D(K/L, \Omega U/L) \rightarrow D(K,\Omega)$, so that the dimensionless counterpart of eq.~(\ref{Dexpr}) is:
\begin{equation}
\begin{aligned}
    D(K,\Omega) = \Omega^2\Omega_{cg}^2 &- \Fr^2(2 C_s^2K^2  - \Omega^2)^2 \\
    & +4 \Fr^2 C_s^3 K^3 \sqrt{C_s^2 K^2-\Omega^2}, 
    \label{Disp}
\end{aligned}
\end{equation}
where $\Omega_{cg}^2 \coleq K + \Bo^{-2} K^3$ is the dimensionless dispersion relation for gravity-capillary waves, where $\Bo \coleq \sqrt{\rho g L^2/\gamma}=L/(\lce \leg)^{1/2}$ is the Bond number, $\Fr=U/\sqrt{gL}$ is the Froude number, and $C_s \coleq c_s/U$ the scaled shear-wave speed in the soft solid. 

The dispersion relation for the soft surface, given by $D(K,\Omega)=0$, interpolates between two venerable classical problems involving surface waves. When $C_s=0$, we recover the well-known dispersion relation for deep water gravity-capillary waves~\cite{landau_lifshitz_fluid_mechanics,lighthill1978waves}, with $\Omega=\Omega_{cg}/\Fr$. In the other limit, when $\Omega_{cg}=0 ~(\Fr \rightarrow \infty)$, the dispersion relation satisfies $\Omega^2= 2 C_s^2K^2 + 2 \sqrt{C_s^3 K^3 \sqrt{C_s^2 K^2-\Omega^2}}$ \footnote{Setting $y=\Omega^2/K^2C_s^2$ we can rewrite this relation as $y(y^3-8y^2+24y-16)=0$, with a single non-trivial real root $y\approx 0.912$, i.e. $\Omega/KC_s \approx 0.955$.} leading to the well-known Rayleigh wave speed on the surface of an incompressible half-space with $v_R = \Omega/K \approx 0.955 C_s$~\cite{rayleigh1885waves}.

In our experiments, with $c_s\approx 0.08~m/s$ and $U\in(0.5, ~1.4)~m/s$, we have $C_s< 0.2$. Assuming that the large-wavenumber contributions are suppressed by the finite bandwidth of the pressure source ($C_s K \ll 1$), the effects of elasticity may be treated perturbatively. This allows us to derive a dispersion relation that minimally captures the leading order effects due to elasticity, but unlike in the 1-dimensional case~\cite{onodera1998surface,shao2018extracting}, the 2-dimensional problem warrants a distinction between modes as a function of the phase speeds $(C_p \coleq c_p/U)$ relative to the shear wave speed $(C_s)$. 

To facilitate discussion of these two different modes, we define a parameter $\lambda(K) \coleq C_s/C_p(K)$. Then, for modes propagating slower than the shear-wave ($\lambda>1$), i.e., $C_p<C_s\ll 1$, the function $D$ (\ref{Disp}) is real-valued and   reduces to:
\begin{equation}\label{eq:slowwaveD}
    D(K,\Omega) = \Omega^2(\Omega_{cg}^2 + 4\varphi_1(\lambda^2) \Fr^2 C_s^2 K^2 - \Fr^2 \Omega^2)~,
\end{equation}
where $\varphi_1:[1,\infty) \rightarrow [0,1/2)$ is a smooth bounded function (see SI B.2). We see that the softness of the elastic half-space introduces a quadratic correction to the dispersion relation for classical gravity-capillary waves on a liquid half-space. In contrast, for modes traveling faster than the shear-wave $(\lambda<1)$, i.e., $C_s < C_p < 1$, the function $D$ (\ref{Disp}) becomes complex-valued and reduces to:
\begin{equation}\label{eq:fastwaveD}
\begin{aligned}
   D(K,\Omega) = \Omega^2(\Omega_{cg}^2 + 4 \varphi_2(\lambda^2) &\Fr^2 C_s^2 K^2 - \Fr^2\Omega^2) \\
   &+ 4i  \Omega \Fr^2 C_s^3 K^3 \sqrt{\varphi_2(\lambda^2)},
   \end{aligned}
\end{equation}
where $\varphi_2:[0,1]\rightarrow [0,1]$ is yet another smooth bounded function (see SI B.2). Here, the softness introduces an additional leading-order cubic contribution to the imaginary part of $D$, and the solution to $D(K,\Omega)=0$ lies in a small neighborhood of the solution to $\Re(D) = 0$ in the complex $K-$plane. Thus, elasticity introduces an effective damping of the propagating surface waves even in the absence of any viscous effects in our energy-conserving theory --  a consequence of bulk radiation that leads to leaky waves~\cite{onodera1998surface}.

\begin{figure}[!t]
\centering
\includegraphics[width=0.85\columnwidth]{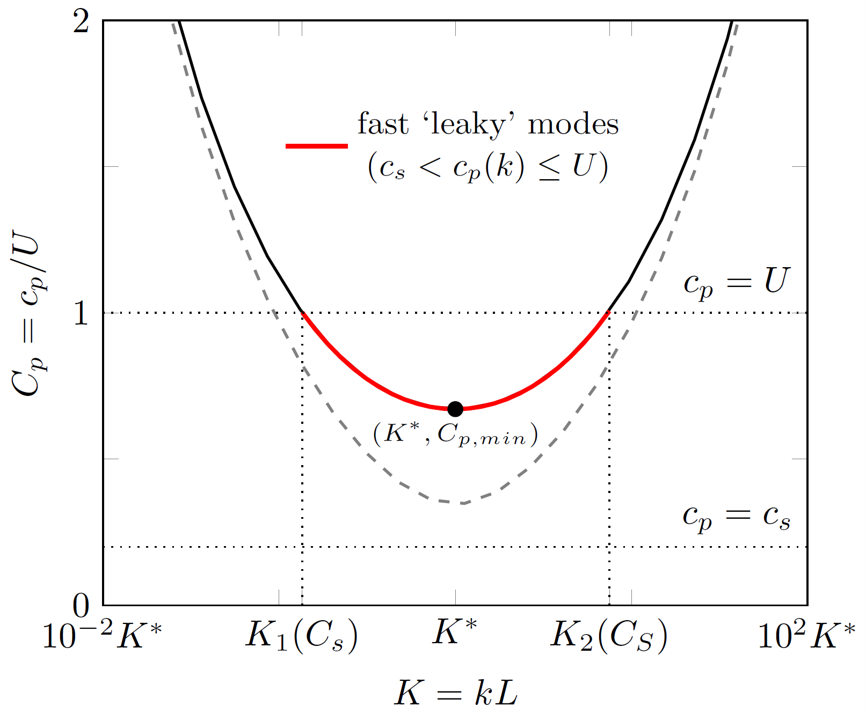}
\caption{Dimensionless phase speed $C_p$ of free waves on the surface of an infinitely deep ultra-soft solid with weak elasticity $(C_s \ll 1)$, shown as a function of the scaled wavenumber $K$ following (\ref{redDis_phasespeed}). The minimum occurs at $K^*=(\rho g L^2/\gamma)^{1/2}$, with a value $C_{p,min} = (2/(\Fr^2 \Bo) + 4 \varphi_2(\lambda^2)C_s^2)^{1/2}$.  The dashed line corresponds to the capillary-gravity waves in infinitely deep inviscid liquid without elasticity $(C_s=0)$.} 
\label{fig:phasespeeds}
\end{figure}

For our experimental conditions, where the quantity $\mu/\sqrt{\rho g \gamma} = \sqrt{\leg/\lce}\sim {0.2}$ is small, there are no real solutions to $D(K,\Omega)=0$ at the critical condition  $\Omega=C_s K$, which precludes the simultaneous existence of fast and slow modes. Furthermore, we find that only the `leaky' fast modes are supported, owing to the very low shear modulus of the ultra-soft gel (see section B.2 in SI)\footnote{For comparison, a solid with much larger modulus, say $\mu \gg 10^2 Pa$, can support both fast and slow modes.}. Henceforth, we consider only the reduced (real-part of the) dispersion relation for fast modes from eq.~(\ref{eq:fastwaveD}) to describe the wake pattern:
\begin{equation}
   \Fr ~\Omega = (\Omega_{cg}^2 + 4 \varphi_2(\lambda^2)\Fr^2  C_s^2K^2)^{1/2} ,\label{red_Disp}
\end{equation}
which yields  the dimensionless phase speed:
\begin{equation}\label{redDis_phasespeed}
    \Fr^2 C_p^2 = 1/K + K/\Bo^2 + 4\varphi_2(\lambda^2)\Fr^2 C_s^2 ~.
\end{equation}
This curve,  plotted in Fig.~\ref{fig:phasespeeds}, shows that the phase speed has a minimum at $K^*=\Bo$, with the corresponding value $C_{p,min} = (2/(\Fr^2 \Bo) + 4  \varphi_2(\lambda^2)C_s^2)^{1/2}$. We see that finite elasticity associated with non-zero $C_s$ introduces a correction to the well-known minimum gravity-capillary phase speed $(2/\Fr^2\Bo)^{1/2}$ \cite{Lamb1932,rabaud2014gravcap}, and leads to a vertical shift of the curve relative to the classical capillary-gravity case, shown in Fig.~\ref{fig:phasespeeds}. However, the minimum still occurs at the same scaled   gravity-capillary wavenumber $K^* =(\rho gL^2/\gamma)^{1/2}$.  

The condition of stationarity that derives from (\ref{eq:CoS}) determines the allowable wavenumbers for pattern formation, so that setting $C_p=\Omega/K=1$ in eq.~(\ref{redDis_phasespeed}) and solving for $K$ gives the critical wavenumbers $K_{1,2}(C_s)$. Consequently, wavenumbers in the range $K \in [K_1(C_s), K_2(C_s)]$ contribute to constructive interference, as shown in Fig.~\ref{fig:phasespeeds}.  For a given pressure source speed, the range of allowable wavenumbers becomes narrower with increasing elasticity, owing to the vertical shift of the curve shown in Fig.~\ref{fig:phasespeeds}. 

\begin{figure}[!b]
\centering
 \includegraphics[trim={0.1cm, 0, 0.1cm, 0.5cm}, clip, width=0.85\columnwidth]{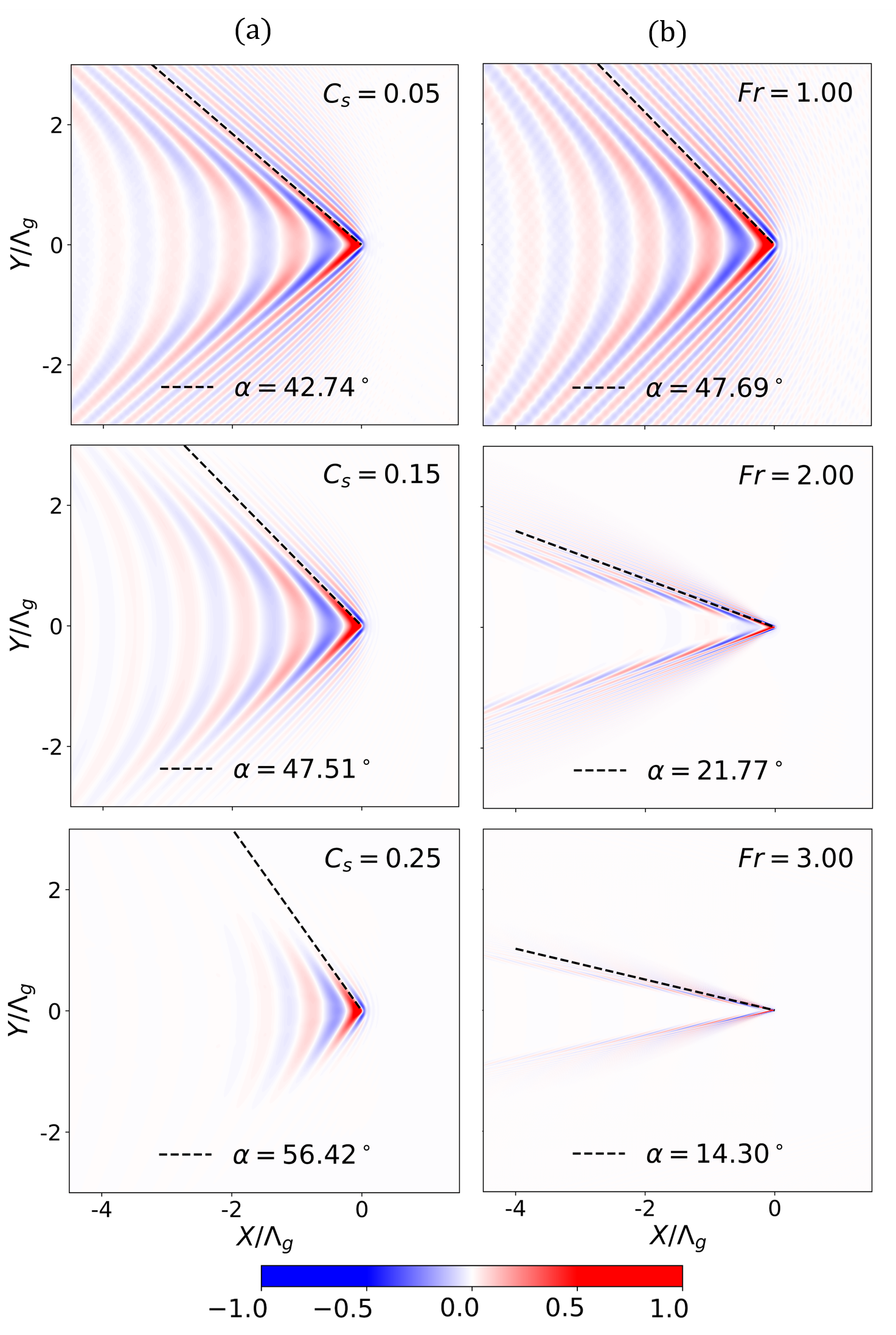}%  
\caption{Wake pattern on ultra-soft solids showing the normalized surface elevation, computed using (\ref{ND-disp}) at $\text{Bo} \approx 3.7$, for different scaled shear-wave speeds $C_s=c_s/U$ and Froude numbers $\Fr$ as functions of the scaled coordinates $X/\Lambda_g$ and $Y/\Lambda_g$, where $\Lambda_g =2\pi \text{Fr}^2$ is the dimensionless wavelength of the gravity wave corresponding the source size $L$. The opening half-angle $\alpha$ of the leading dominant ridge (dashed black line) is shown for increasing (a) $C_s$ at $\Fr=1.1$ and (b) $\Fr$ at $C_s=0.02$. The computational domain is approximately $0.75~ \Fr^2$ times the length of the experimental tank.}%
\label{fig3:simulations}%
\end{figure}

\begin{figure}[!t]
\includegraphics[width=.9\columnwidth]{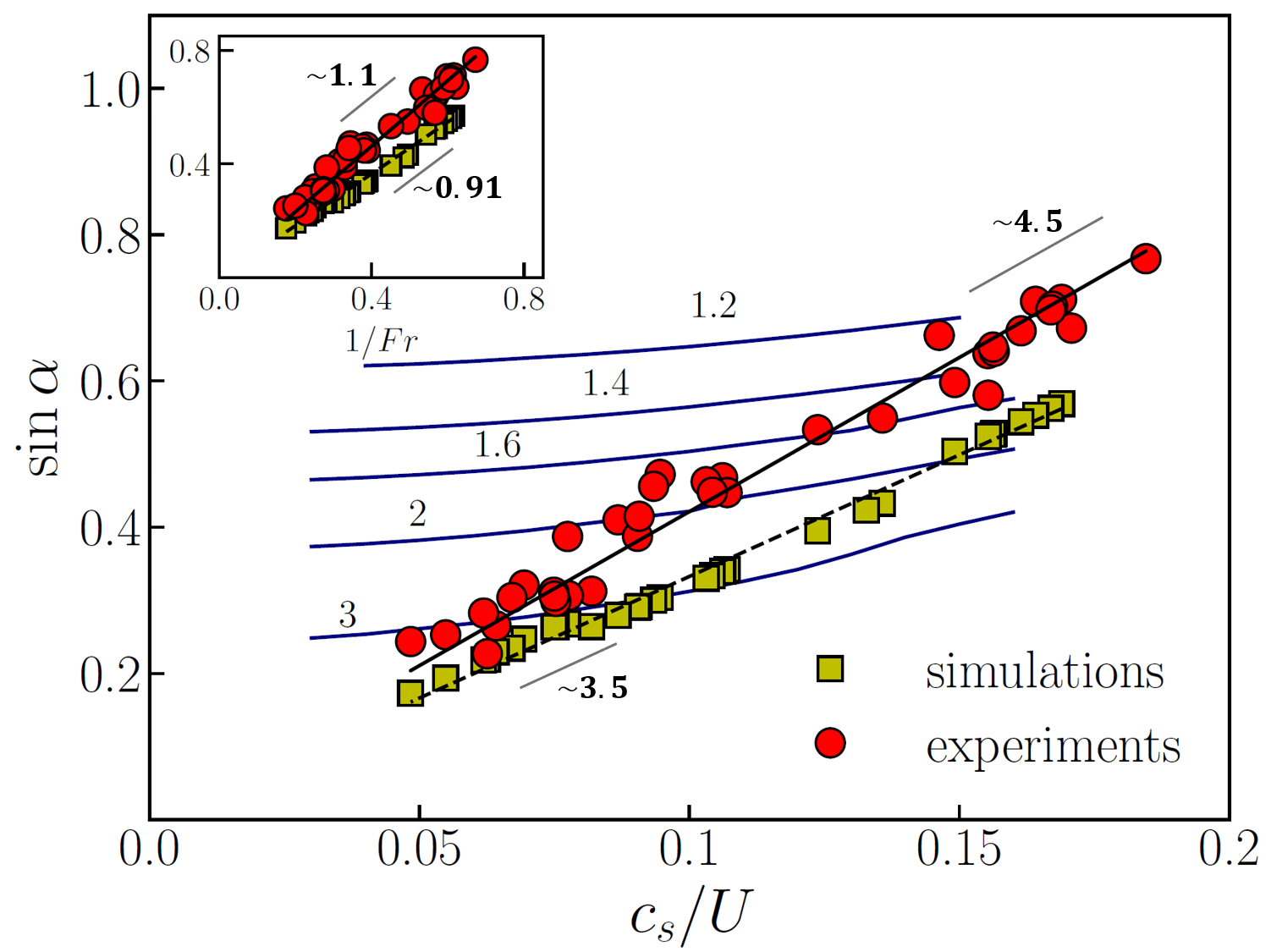}
    \caption{Plot of $\sin \alpha$ as a function of $c_s/U$ for different Froude numbers $\Fr$, obtained from simulations following (\ref{ND-disp}). Blue curves: iso-$\Fr$ contours from simulations, with $\Fr$ indicated above each curve. Yellow squares: simulations run at the experimental conditions. Red circles: experimental observations for comparison. The experimental data points that correspond to different pulse speeds $U$ move across iso-$\Fr$ curves, such that $\Fr ~C_s \sqrt{\Bo} = \frac{U}{\sqrt{gL}}\sqrt{\frac{\mu}{\rho U^2}}(\frac{\rho gL^2}{\gamma})^{1/4} =[(\mu^2/(\rho g \gamma)]^{1/4}$ is constant. The inset shows the wake angle scales with the inverse of $\Fr$, for both experiments and simulations.}
    \label{fig:anglecomparison}
\end{figure}
    
To compare our theory with observations, we perform numerical simulations to  compute the far-field surface wave amplitude that develops in response to an applied pressure field using eq.~(\ref{eq:surface_elev_integral}), with the complete dispersion expression corresponding to eq.~(\ref{Disp}). We model the pressure source  as a radially-symmetric Gaussian field of size $L$  moving with uniform velocity $\bm{U}$ in the $x-$direction: $ P_z(\bm{x},t; U,L) = P_0\exp{[-\pi^2(|\bm{x}-\bm{U}t|)^2/L^2]}.$ Introducing dimensionless variables $\bm{X} \coleq \bm{x}/L, ~W \coleq w/L$, and recalling the earlier scaling $ K=k L$, ~$\Omega=\omega L/U$, the scaled version of the surface elevation in the co-moving frame in eq.(\ref{eq:surface_elev_integral}) is: 
\begin{equation}
W(\bm{X}) = \lim_{\epsilon \rightarrow 0}\frac{1}{(2\pi)^2}\Re \int \int ~d^2\bm{K}~  e^{i\bm{K} \cdot \bm{X}} \frac{\Omega_{\epsilon}^2 K \tilde{P}_z^*(K)}{D(K,\Omega_{\epsilon})} ,\label{ND-disp}
\end{equation}
where $D$ follows from eq.~(\ref{Disp}),
$\Omega_{\epsilon} \coleq \Omega + i\epsilon = \hat{\bm{U}} \cdot \bm{K} + i \epsilon~,$ and  $\tilde{P}^*_z(K) ~ \coleq \tilde{P}_z(K)/(\rho g L^3) \propto \exp{(-K^2/4\pi^2)}$. For the assumed pressure field, we numerically compute $W(\bm{X})$ using a 2-dimensional inverse fast Fourier transform on a fixed square domain (spanning $6$ wavelengths of gravity wave), discretized on an $N\times N=4096^2$ grid, with the regularization $\Omega \rightarrow \Omega_{\epsilon}$, choosing $\epsilon \approx 10^{-2}$ as a compromise between artificial oscillations introduced by small $\epsilon$ and exponential decay at large $\epsilon$, respectively (see SI for details). 

Sweeping over the scaled parameter space $(\Fr, C_s)$, keeping $\Bo$ fixed allows us to compare the results with our experiments. In Fig.~\ref{fig3:simulations}, we show the wake patterns characterized by the opening semi-angle $\alpha$ of the leading ridge \footnote{The half-angle $\alpha$ defines the leading dominant ridge, which was visible and measured in the experiments; this is distinct from the wake angle based on $\beta$ given by (\ref{eq:wake-angle-of-K}), but that was hard to observe.} (see Fig.~\ref{fig:diffAngles} in End Matter for definitions of $\alpha$ and standard wake angle), from two different sets of simulations performed with $\Bo \approx 3.7$. In Fig.~\ref{fig3:simulations}a, the scaled shear-wave speed $C_s$ is varied while keeping the Froude number constant. The opening semi-angle $\alpha$  increases with increasing elasticity, and the wake distinctly becomes more spatially damped, consistent with both the theoretical predictions and the experimental trend as shown in Fig.~\ref{experimentandData}b, c. In Fig.~\ref{fig3:simulations}b, we show the effects of the Froude number $\Fr$ on wake angle, that can be achieved by changing either the disturbance size $L$ or speed $U$. The opening angle decreases at large $\Fr$, in qualitative agreement with our experimental observations in Fig.~\ref{experimentandData}b.

To quantitatively compare the theory with our experiments, we measure wake opening angles $\alpha$ from simulations for $\Fr \in [1, 3]$ and $C_s \in [0.01, 0.2]$. In Fig.~\ref{fig:anglecomparison} we see that the linear relation obtained using geometric considerations persists, with $\sin\alpha = A c_s/U$; simulations yield $A\approx 3.5$, while experiments yield $A \approx 4.5$. Similarly, the dependence on $\Fr$ follows a Mach-like law, $\sin\alpha =B/\Fr$, with $B\approx 0.91$ from simulations and $B \approx 1.1$ from experiments, showing reasonable agreement. We believe that the primary reason for the difference is finite size effects associated with not being strictly in the deep-water limit.

Our study of surface wakes on ultra-soft elastic surfaces uses experiments and theory to probe a previously unexplored regime where gravity, capillarity, and elastodynamics act together, while bringing together the classical subjects of Rayleigh waves and Kelvin wakes.  Our experiments allow us to observe wakes behind a moving source as a function of its speed and our theory shows how the geometric simplicity of Mach-like wakes combines with the dispersive character of gravity–capillary waves. Furthermore, we see that elasticity leads to effective damping through radiative leakage, even in the absence of viscosity. All together, our approach also provides a quantitative foundation for probing the dynamics of ultra-compliant solid surfaces.

{\bf Acknowledgments.} This work was supported in part by the Simons Foundation and the Henri Seydoux Fund. The computations in this paper were run on the FASRC FASSE cluster supported by the FAS Division of Science Research Computing Group at Harvard University.

\bibliographystyle{apsrev4-1}
%\bibliography{surfacewakes}

%

\newpage
\begin{center}
    \textbf{END MATTER}
\end{center}
\begin{center}
\textbf{Radiation and Wake Angles}
\end{center}

To characterize the nature and form of the wake, we examine the group speed of the wave packets generated by the dispersive system (\ref{red_Disp}), since it determines where the energy of the generated waves focuses spatially. The scaled group speed, $C_g = c_g/U$, corresponding to eq.~(\ref{red_Disp}) is:
\begin{equation}\label{eq:group_speed}
    C_g(K):= \frac{\partial \Omega}{\partial K} =  \frac{1 + 3K^2/\Bo^2 + 8\varphi_2(\lambda^2) \Fr^2 C_s^2 K}{2 \Fr^2\Omega}~.
\end{equation}
In Fig.~\ref{fig:EM-A}a, we plot $C_g$ and the phase speed $C_p=\Omega/K$ as a function of $K$. Just as for classical gravity-capillary waves, there are two distinct branches: waves with $K\in[K_1(C_s), K^*]$ lie on the \textit{elasto-gravity branch}, where $C_g < C_p$, characteristic of   gravity waves, and those with $K \in [K^*, K_2(C_s)]$ lie on the \textit{elasto-capillary branch}, where $C_g>C_p$ analogous to  capillary waves.
\begin{figure}[!h]
    \centering
\includegraphics[width=\columnwidth]{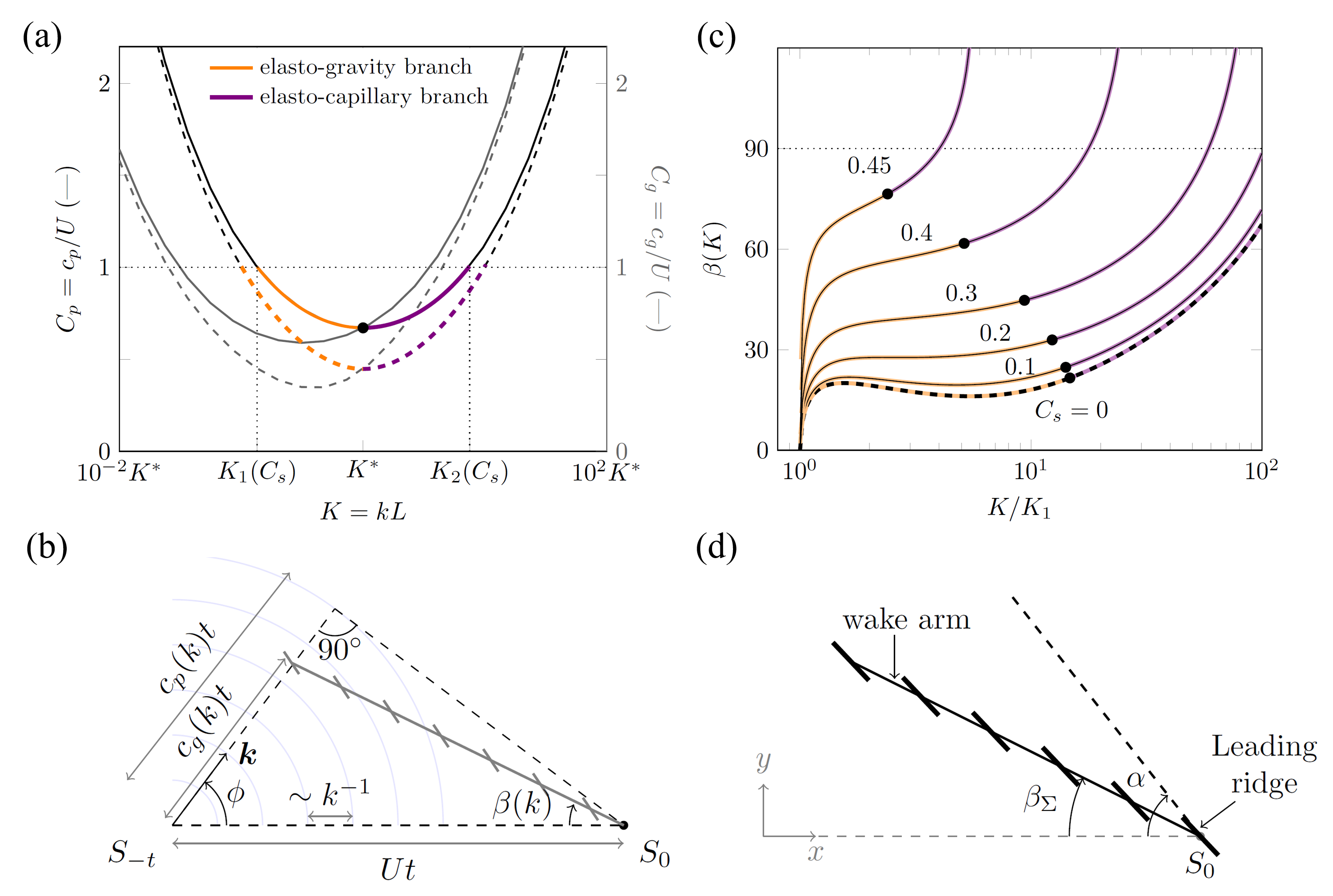}  
    \caption{(a) Dimensionless phase and group speeds as functions of wavenumber from gravity-capillary elastodynamics ($C_s>0)$ following (\ref{redDis_phasespeed}) and (\ref{eq:group_speed}), showing the \textit{elasto-gravity} ($C_g<C_p$) and \textit{elasto-capillary} branches  ($C_g>C_p$). (b) Schematic of the stationary wake pattern from interference of monochromatic waves generated by a source moving from $S_{-t}$ to $S_0$, adapted from ~\cite{crawford1984elementary,rabaud2014gravcap}.  (c) Radiation angle $\beta$ for allowed wavenumbers at fixed $\Bo, ~\Fr$, with varying shear-wave speeds $C_s$, following (\ref{eq:wake-angle-of-K}). The dashed curves in (b) and (c) correspond to $C_s=0$. (d) Schematic of the resultant wake pattern formed by superposition of broadband waves, defining the characteristic angles.}
    \label{fig:EM-A}
\end{figure}

 In Fig.~\ref{fig:EM-A}b, we show the geometry of monochromatic waves associated with a single wavenumber generated by the moving source. As it moves from $S_{-t}$ to $S_0$, where $t$ denotes the elapsed time, circular wave fronts emanate from successive source locations on the horizontal axis. Waves generated earliest at $-t$ travel the farthest at the phase speed $c_p$, and their constructive interference with the waves generated subsequently produces focusing of wave amplitudes along a particular line. For each admissible wavenumber in the range $[K_1(C_s), K_2(C_s)]$, this line is characterized by the radiation angle $\beta(K)$ between the source velocity vector $(-\bm{U})$ and the group velocity vector of the wave packet in the moving frame ($c_g \bm{\hat{k}} - \bm{U}$). Following the geometric constructions in \cite{crawford1984elementary,rabaud2014gravcap}, and considering the magnitudes and directions of the stationary phase \cite{thomson1887ship}, the radiation angle of the wave group of a given $K$ is:  
\begin{equation}\label{eq:wake-angle-of-K}
    \tan\beta(K) = \frac{\Cg(K)\sqrt{1-\Cph^2(K)}}{1-\Cg(K)\Cph(K)}~.
\end{equation}

We plot $\beta(K)$ in Fig.~\ref{fig:EM-A}c for different shear-wave speeds by varying $C_s$. Unlike pure gravity waves in deep water where the wake arm is bounded at the classical Kelvin angle, the radiation angle of free waves on ultra-soft solid spans all directions, similar to gravity-capillary waves. Wavenumbers on the elasto-gravity branch radiate behind source ($\beta<90^o$), whereas those on the elasto-capillary branch radiate in arbitrary directions including ahead of the source ($\beta>90^o$). Increasing elasticity increases the radiation angle of every allowable wavenumber. However, the larger wavenumbers, which radiate at larger angles, are strongly damped due to elasticity, and thus appear less prominently in the wake pattern.

The constructive interference of the individual wave packets of different wavenumbers, each radiating in different angles $\beta(K)$, generates the wake pattern with multiple ridges behind the source (see Fig.~\ref{fig:EM-A}d). The elevation peaks along these ridges determine the observed wake angle $\beta_{\Sigma}$, which is different from the opening angle of the leading dominant ridge $\alpha$. This distinction is shown for a representative set of parameters in Fig.~\ref{fig:diffAngles}. We note that spatial damping due to elasticity makes the observation and measurement of $\beta_{\Sigma}$ in the experiments increasingly difficult for larger $C_s$, thus driving our choice to measure $\alpha$ rather than the entire wake pattern.
\begin{figure}[!h]
    \centering
     \includegraphics[trim={0, 0, 0, 0}, clip, width=\columnwidth]{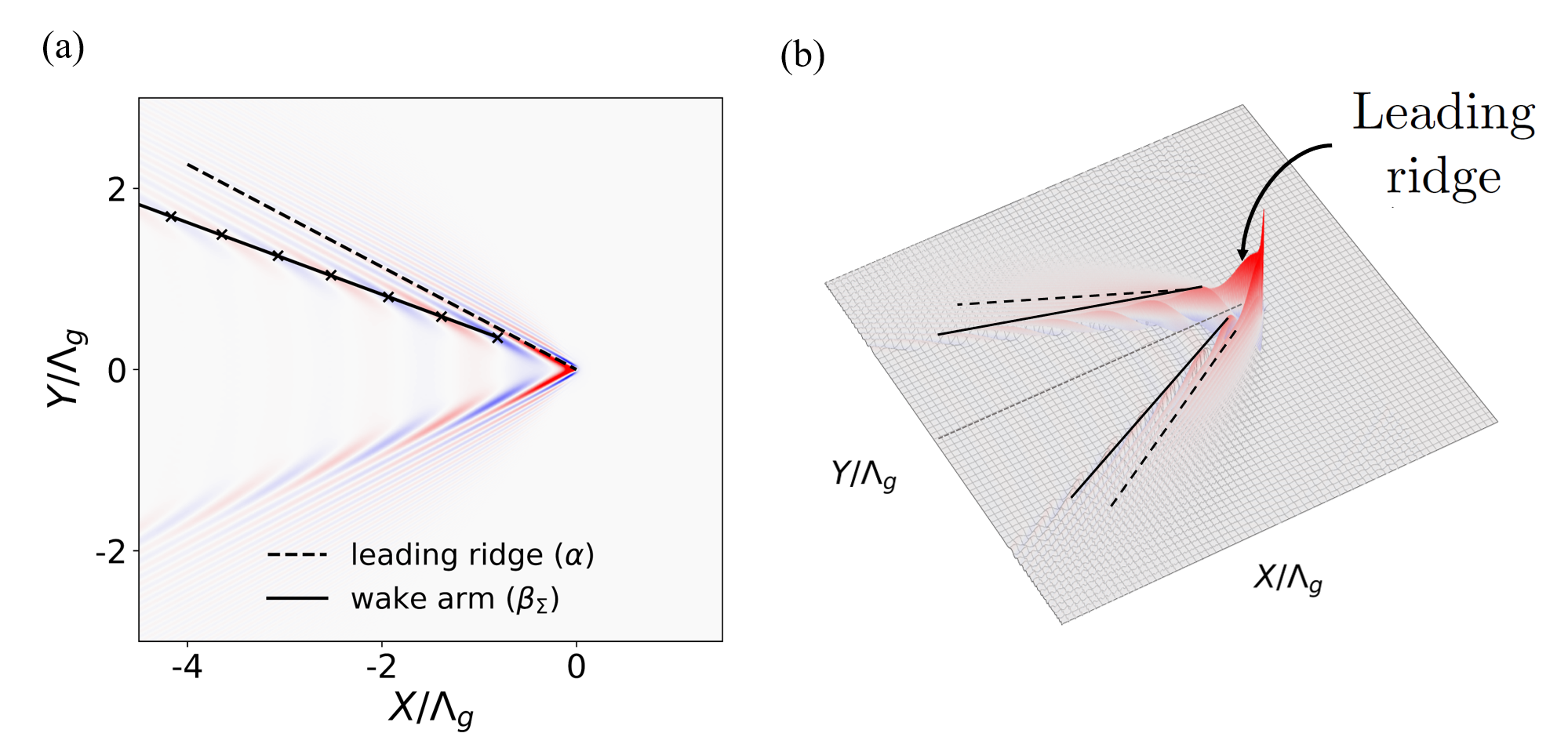}
    \caption{(a) Top and (b) 3D views of the surface wake pattern for a representative case ($\Fr=1.5, ~\Bo=3.7,~ C_s=0.01$), showing the difference between the leading dominant ridge (dashed), characterized by the opening angle $\alpha$, and the wake arm (solid), characterized by the wake angle $\beta_{\Sigma}$. The markers ($\times$) in (a) indicate the local extremum of the surface displacement field in \textit{each} ridge, and the connecting line passes through these extrema across the successive ridges. }
 \label{fig:diffAngles}
\end{figure}

\end{document}